\documentstyle[12pt]{article}
\def\fun#1#2{\lower3.6pt\vbox{\baselineskip0pt\lineskip.9pt
  \ialign{$\mathsurround=0pt#1\hfil##\hfil$\crcr#2\crcr\sim\crcr}}}
\newskip\humongous \humongous=0pt plus 1000pt minus 1000pt

\newif\ifdtup

\def\oldreffmt#1{\rlap{[#1]} \hbox to 2\parindent{}}

\def\figfmt#1{\rlap{Figure {#1}} \hbox to 1in{}}

\def\beq{\begin{equation}}
\def\eeq{\end{equation}}

\def\bq{\begin{quote}}
\def\eq{\end{quote}}

\relax
\addtocounter{equation}{-1}
\newcommand{\be}{\begin{equation}}
\newcommand{\ee}{\end{equation}}
\begin{document}

\begin{titlepage}
\rightline{ETH-TH/92-21}
\rightline{CPT-92/P.2708}
\vskip 0.5truecm
\begin{center}
{\large{\bf Lie-algebraic approach to the theory of polynomial solutions.\\
II. Differential equations in one real and one Grassmann variables\\
and\\ 2x2 matrix differential equations}}
\footnote{This work was supported in part by the Swiss National Science
Foundation}
\vskip 0.5truecm
 {\bf A.Turbiner}\footnote
{On leave of absence from: Institute for Theoretical and Experimental Physics,
Moscow 117259, Russia\\E-mail: turbiner@cernvm or turbiner@vxcern.cern.ch}
\vskip 0.5cm
Theoretische Physik, ETH-Honggerberg, CH-8093 Zurich, Switzerland\\ and\\
Centre de Physique Theorique, Marseille Cedex 9, F-13288, France
\\
(submitted to Comm.Math.Phys.)
\end{center}
\vskip 0.5truecm
\begin{center}
{\large ABSTRACT}
\end{center}
\begin{quote}

A classification theorem for linear differential equations
in two variables (one real and one Grassmann) having polynomial
solutions(the generalized Bochner problem) is given. The main result is
based on the consideration of the eigenvalue problem for a polynomial
 element of the universal enveloping algebra of the algebra $osp(2,2)$
 in the "projectivized" representation (in differential operators
 of the first order) possessing an  invariant subspace. A classification
of 2 x 2 matrix differential equations in one real variable possessing
 polynomial solutions is described. Connection to the recently-discovered
quasi-exactly-solvable problems is discussed.
\end{quote}

\vfill

\end{titlepage}
\newpage
\baselineskip=18pt plus 1pt minus 1pt

Take the eigenvalue problem
\be
 T \varphi  \  = \ \epsilon \varphi
\ee
where $T$ is a linear differential operator of one real $x \in {\bf R}$
and one Grassmann $\theta$ variables and $\epsilon$ is the spectral parameter.

{\bf Definition.} Let us give the name {\it generalized Bochner
problem} to the problem of classification of the differential equations
(0) having $(2n+1)$ eigenfunctions in the form of polynomials in $x,\theta$
of a degree not higher than $n$.

In Ref.\cite{t1} a general method  has been formulated for generating
 eigenvalue problems for linear differential
 operators, linear matrix differential operators and
 linear finite-difference operators in one and several variables
possessing polynomial solutions. The
method is based on considering the eigenvalue problem for the representation
of a polynomial element of the universal enveloping algebra of the Lie
algebra in a finite-dimensional, 'projectivized'
representation of this Lie algebra \cite{t1}.

In a previous paper \cite{t2} it has been proven that in this approach
consideration of the algebras $sl_2(\bf R)$ and $sl_2(\bf R)_q$ in
projectivized representations provides both necessary and sufficient
conditions for existence
of polynomial solutions in ordinary linear finite-order differential equations
and in a certain class of finite-difference  equations in one variable,
respectively. Particularly, it manifested the classification theorems,
which imply the solution of the Bochner problem (1929) posed for ordinary
differential equations. In the present paper a similar classification
theorem is given for finite-order linear differential equations
in two variables: one real and one Grassmann, in connection to the algebra
$osp(2,2)$. Also presented is a consideration of 2 x 2
matrix differential equations in one real variable, which is closely
connected to the previous problem of one real and one Grassmann variables.

\section{Generalities}

\hspace{13mm}
Define the following space of polynomials in $x,\theta$
\be
\label{e1}
{\cal P}_{N,M} \ = \ \langle x^0,x^1,\dots,x^N, x^0\theta, x^1\theta, \dots,
x^M \theta \rangle
\ee
where $N, M$ are non-negative integers, $x \in {\bf R}$ and $\theta$ is
Grassmann (anticommuting) variable. \par

The projectivized representation of the algebra $osp(2,2)$ is given as follows.

The algebra  $osp(2,2)$ is characterized by four bosonic generators
$T^{\pm ,0},J$ and four fermionic generators $Q_{1,2},\overline{Q}_{1,2}$
and given by the commutation and anti-commutation relations
\label{e2}
\[
[T^0 , T^{\pm}]= \pm T^{\pm} \quad , \quad [T^+ , T^-]= -2 T^0 \quad , \quad
 [J , T^{\alpha}]=0  \quad , \alpha = +,-,0 \]
\[   \{ Q_1,\overline{Q}_2 \}  = - T^- \quad , \quad \{ Q_2, \overline{Q}_1 \}
= T^+ \quad, \]
\[ {1 \over 2} \{ \overline{Q}_1, Q_1 + \overline{Q}_2, Q_2 \} = - J \quad ,
\quad
 {1 \over 2} \{ \overline{Q}_1, Q_1 - \overline{Q}_2, Q_2 \} =  T^0 \quad , \]
\[  [Q_1 , T^+]=Q_2 \quad , \quad [Q_2 , T^+]=0 \quad , \quad [Q_1 ,
T^-]=0 \quad , \quad [Q_2 , T^-]=-Q_1 \ ,\]
\[  [\overline{Q}_1 , T^+]=0 \quad , \quad  [\overline{Q}_2 , T^+] =
\overline{Q}_1 \quad , \quad
[\overline{Q}_1 ,T^-] = \overline{Q}_2 \quad , \quad [\overline{Q}_2 ,
T^-]=0 \ ,\]
\[    [Q_{1,2} , T^0]=\pm {1 \over 2} Q_{1,2} \quad , \quad
 [\overline{Q}_{1,2} , T^0]=\mp {1 \over 2} \overline{Q}_{1,2} \]
\be
  [Q_{1,2}, J]=\pm {1 \over 2} Q_{1,2} \quad , \quad
 [ \overline{Q}_{1,2}, J]=\pm {1 \over 2} \overline{Q}_{1,2}
\ee
This algebra possesses the projectivized representation \cite{st}
\label{e3}
\[ T^+ = x^2 \partial_x - n x + x \theta \partial_{\theta},  \]
\be
 T^0 = x \partial_x - {n \over 2} +{1 \over 2} \theta \partial _{\theta }\  ,
\ee
\[ T^- = \partial_x \ .  \]
\[ J = -{ n \over 2} - {1 \over 2} \theta \partial_{\theta} \]
for bosonic (even) generators and
\be
\label{e4}
Q= { Q_1 \brack Q_2 } = { \partial_{\theta} \brack x\partial_{\theta}} ,
{\bar Q} = { {\bar Q}_1 \brack {\bar Q}_2 } =
{ x\theta\partial_x-n \theta \brack -\theta\partial_x} ,
\ee
for fermionic (odd) generators, where $x$ is a real variable and $\theta$ is
a Grassmann variable. Inspection of the generators shows that if $n$ is a
non-negative integer, the representation Eqs.(3),(4) is finite-dimensional
representation of dimension $(2n+1)$. The polynomial space ${\cal P}_{n,n-1}$
describes the corresponding invariant sub-space.

{\bf Definition.} Let us name a linear differential operator of the $k$-th
order, ${\bf T}_k (x,\theta)$, {\bf quasi-exactly-solvable} if it preserves
 the space ${\cal P}_{n,n-1}$. Correspondingly, the operator
${\bf E}_k(x,\theta) \in {\bf T}_k(x,\theta)$, which preserves
the infinite flag $ {\cal P}_{0,0} \subset  {\cal P}_{1,0}
\subset {\cal P}_{2,1}
\subset \dots \subset {\cal P}_{n,n-1} \subset \dots $ of spaces of all
polynomials, is named {\bf exactly-solvable}.

{\bf LEMMA 1. } {\it Take the space ${\cal P}_{n,n-1}$.

 (i) Suppose $n > (k-1)$.  Any quasi-exactly-solvable operator of
$k$-th order ${\bf T}_k (x,\theta)$, can be
represented by a $k$-th degree polynomial of the operators (3),(4).
 If $n \leq (k-1)$, the part of the quasi-exactly-solvable operator
${\bf T}_k (x,\theta)$ containing
derivatives in $x$ up to order $n$ can be represented by an $n$-th
degree polynomial in the generators (3), (4).

(ii) Inversely, any polynomial in (3),(4) is quasi-exactly solvable
operator.

(iii) Among quasi-exactly-solvable operators
there exist exactly-solvable operators ${\bf E}_k \subset {\bf T}_k (x,
\theta)$.}\par

{\it Comment 1.} If we define the universal enveloping algebra $U_g$ of a
 Lie algebra $g$ as the algebra of all polynomials over the generators,
then the meaning of the Lemma is the following:
${\bf T}_k (x,\theta)$ at $k < n+1$
is simply an element of the universal enveloping algebra $U_{osp(2,2)}$
of the algebra $osp(2,2)$ in the representation (3),(4).
If $k \geq n+1$, then  ${\bf T}_k (x,\theta)$ is
represented as a polynomial in (3),(4) of degree $n$ plus $B {\partial ^{n+1}
\over {\partial x^{n-m+1} \partial \theta^{m}}}$  , where $m=0,1$ and
 $B$ is any linear differential operator of order not higher than
$(k-n-1)$. \par

{\bf Proof.} The proof is technical and but a straightforward analogue of the
proof of the similar lemma for the case of linear differential operators in
one variable (see Lemma 1 in \cite{t2} and its proof)
\footnote{Recently, J.Frohlich and M.Shubin suggested more simple proof, based
on irreducibility of the representation ${\cal P}_{N,N-1}$ and further use of
the Burnside theorem (see e.g.\cite{w}).}.

Let us introduce the grading of the bosonic generators (3)
\be
\label{e5}
deg (T^+) = +1 \ , \ deg (J,T^0) = 0 \ , \ deg (J^-) = -1
\ee
and fermionic generators (4)
\be
\label{e6}
deg (Q_2,\overline{Q}_1) =+ {1 \over 2}\ , \
deg (Q_1,\overline{Q}_2) =- {1 \over 2}
\ee
Hence the grading of monomials of the generators (3),(4) is equal to
\label{e7}
\[ deg [(T^+)^{n_+} (T^0)^{n_0}J^{\overline{n}}(T^-)^{n_-}{Q_1}^{m_1}
{Q_2}^{m_2}
{\overline{Q}_1}^{\overline{m}_1}{\overline{Q}_2}^ {\overline{m}_2} ]
 \  = \]
\be
  (n_+ - n_-) \ - \ (m_1 - m_2 - {\overline{m}_1} + {\overline{m}_2}) / 2
\ee
The $n$'s can be arbitrary  non-negative integers, while the $m$'s are
equal either 0 or 1. The notion of grading allows us to classify the
operators ${\bf T}_k (x,\theta)$ in a Lie-algebraic sense.\par

{\bf LEMMA 2. } {\it A quasi-exactly-solvable operator
${\bf T}_k (x,\theta) \subset U_{osp(2,2)}$
 has no terms of positive grading other than  monomials of
grading +1/2 containing the generator $Q_1$ or $Q_2$,
 iff it is an exactly-solvable operator.} \par

{\bf THEOREM 1.} {\it Let $n$ is a non-negative integer. Take the eigenvalue
problem for a linear differential
operator in one real and one Grassmann variables
\be
\label{e8}
 {\bf T}_k (x,\theta) \varphi \ = \ \varepsilon \varphi
\ee
\noindent
where $ {\bf T}_k$ is symmetric in a certain manner. In general,
the problem (8) has $(2n+1)$ linear independent eigenfunctions
in the form of polynomials in variable $x,\theta$ of order
not higher than $n$, iff $T_k$ is quasi-exactly-solvable. The problem
(8) has an infinite sequence of polynomial eigenfunctions, iff the
operator is exactly-solvable.} \par

This theorem gives a general classification of differential equations
\be
\label{e9}
 \sum_{i,j=0}^{i=k,j=1} a_{i,j} (x,\theta) \varphi^{(i,j)}_{x,\theta}
(x,\theta) \ = \ \varepsilon \varphi(x,\theta) \ ,
\ee
 where the notation $ \varphi^{(i,j)}_{x,\theta}$ means the
$i-$th order derivative with respect to $x$ and $j-$th order derivative with
respect to $\theta$,
having at least one polynomial solution in $x,\theta$, thus resolving the
generalized Bochner problem.
Suppose that $k>0$, then the coefficient functions $a_{i,j} (x,\theta)$
should have the form
\[ a_{i,0} (x,\theta) \ =  \ \sum_{p=0}^{k+i} a_{i,0,p}\ x^p\ +\  \theta
\sum_{p=0}^{k+i-1}\overline{a}_{i,0,p}\ x^p \]
\be
\label{e10}
a_{i,1} (x,\theta) \ = \ \sum_{p=0}^{k+i-1} a_{i,1,p}\ x^p\ + \ \theta
\sum_{p=0}^{k+i-1}
 \overline{a}_{i,1,p}\ x^p
\ee

The explicit expressions (10) are obtained by substituting (3),(4) into
a general, the $k$-th order, polynomial element of the universal
enveloping algebra $U_{osp(2,2)}$. Thus the coefficients $a_{i,j,p}$
can be expressed through the coefficients of the $k$-th order polynomial
element of universal enveloping algebra $U_{osp(2,2)}$. The number of free
parameters of the polynomial solutions is defined by the number of
parameters characterizing  general, a $k$-th order polynomial element of the
universal enveloping algebra $U_{osp(2,2)}$. However, in counting parameters
a certain ordering of the generators  should be introduced to avoid double
counting due to commutation and anti-commutation relations. Also  some
relations between generators should be taken into account,
specific for the given representation (3),(4), like

\[ 2 T^+ J\ -\ \overline{Q}_1 Q_2\ =\ n T^+ \ , \]
\[ T^+ Q_1\ -\ T^0 Q_2\ =\ - Q_2 \ , \]
\[ T^+ \overline{Q}_2\ +\ T^0 \overline{Q} _1\ =\ {(1-n) \over 2}\
\overline{Q}_1 \ , \]
\[ J Q_2\ =\ {n \over 2}\ Q_2 \ ,\]
\[ J \overline{Q}_1\ =\ {(n+1) \over 2}\ \overline{Q}_1 \ ,\]
\[ T^+T^-\ -\ T^0 T^0\ -\ J J\ +\ T^0\ =\ - {n \over 2}(n-1) \ , \]
\[ J J\ =\ (n + {1 \over 2}) J\ -\ { n \over 4 }(n+1) \ , \]
\[ Q_1 \overline{Q}_1\ +\  Q_2 \overline{Q}_2\ -\ 2n J\ =\ - n(n+1) \ , \]
\[2 T^0 J\ +\ Q_1 \overline{Q}_1\ -\ (n+1) T^0\ -\ nJ\ =
\ - {n \over 2} (n+1) \ , \]
\[ T^- Q_2\ -\ T^0 Q_1\ =\  ( {n \over 2} +1) Q_1 \ , \]
\[ T^- \overline{Q}_1\ -\ T^0 \overline{Q}_2\ =\  {(n - 1) \over 2})
\overline{Q}_2 \ , \]
\[ J Q_1\ =\ {n \over 2} Q_1 \ ,\]
\[ J \overline{Q}_2\ =\ {n+1 \over 2} \overline{Q}_2 \ ,\]
\be
\label{e11}
 2 J T^-\ -\ Q_1 \overline{Q}_2\ =\ (n+1) T^-
\ee
between quadratic expressions in generators and the double-sided
ideals generated by them. Straightforward analysis leads to the following
expression for the number of free parameters of a quasi-exactly-solvable
operator ${\bf T}_k (x,\theta)$
\be
\label{e12}
par({\bf T}_k (x,\theta))= 4k(k+1)+1 \ ,
\ee
where we denoted the number of free parameter of the operator $T$ through the
symbol $par(T)$ .  Note, that for the second-order quasi-exactly solvable
operator $par({\bf T}_2)=25$ \footnote{Recall, that for the case of
the second-order differential operator in one
real variable, the number of free parameter  was equal to 9 (see\cite{t2}).}.
For the case of an exactly-solvable operator (an infinite sequence of
polynomial solutions of Eq. (9)), the Eq. (10)
simplifies and reduces to

\[ a_{i,0} (x,\theta) \ =  \ \sum_{p=0}^{i} a_{i,0,p} x^p +  \theta
\sum_{p=0}^{i-1}\overline{a}_{i,0,p} x^p \]

\be
\label{e13}
a_{i,1} (x,\theta) \ = \ \sum_{p=0}^{i} a_{i,1,p} x^p +
 \theta \sum_{p=0}^{i-1} \overline{a}_{i,1,p} x^p
\ee
Correspondingly, the number of free parameters reduces to
\be
\label{e14}
par({\bf E}_k (x,\theta))= 2k(k+2)+1
\ee
For the second-order exactly solvable
operator $par({\bf E}_2(x,\theta))=17$ \footnote{Recall, that for the case
of the second-order differential operator in one
real variable, the number of free parameter  was equal to 6 (see \cite{t2}).}.
Hence, Eq.(9) with coefficient functions (13) gives a general form
of eigenvalue problem for the operator ${\bf T}_k$, which can lead to an
infinite family of orthogonal polynomials as eigenfunctions. If we put
formally all coefficients in (13), $\overline{a}_{i,0,p}$ and $a_{i,1}
(x,\theta)$ equal to zero, we reproduce the eigenvalue problem for the
differential operators in one real variable, which gives rise to all known
families of orthogonal polynomials in one real variable (see \cite{t2}).

\section{Second-order differential equations in $x$,$\theta$}

\hspace{13mm}
Now let us consider in more detail the second order differential equation
Eq.(9), which can possess polynomial solutions. As follows from Theorem 1,
the corresponding differential operator ${\bf T}_2(x,\theta)$ should be
quasi-exactly-solvable. Hence, this operator can
be expressed in terms of $osp(2,2)$ generators taking into account the
relations (11)
\label{15}
\[ {\bf T}_2 =  c_{++} T^+ T^+ + c_{+0} T^+ T^0 +  c_{+-} T^+ T^- +
 c_{0-} T^0 T^- + c_{--} T^- T^- + \]
\[ c_{+J} T^+ J + c_{0J} T^0 J + c_{-J} T^- J + \]
\[ c_{+ \overline{1}} T^+ \overline{Q}_1 +  c_{+2} T^+ Q_2 +
  c_{+1} T^+ Q_1 + c_{+ \overline{2}} T^+ \overline{Q}_2 +
   c_{01} T^0 Q_1 + \]
\[ c_{0 \overline{2}} T^0 \overline{Q}_2 +  c_{-1} T^- Q_1 +
c_{- \overline{2}} T^- \overline{Q}_2 + \]
\be
c_+ T^+ + c_0 T^0 + c_- T^- + c_J J + c_1 Q^1 + c_2 Q^2 +  c_{\overline{1}}
\overline{Q}_1 + c_{\overline{2}} \overline{Q}_2 + c
\ee
where $c_{\alpha \beta}, c_{\alpha}, c $ are parameters.
Following Lemma 2, under the conditions
\be
\label{e16}
 c_{++}  = c_{+0} = c_{+ \overline{1}}= c_{+ \overline{2}}=
c_{ \overline{1}} = c_{+2}  = c_{+J}= c_{+}  =0 \ ,
\ee
the operator ${\bf T}_2(x,\theta)$ in representation (13) becomes
exactly-solvable.

Now we proceed to the detailed analysis of the quasi-exactly-solvable
operator ${\bf T}_2(x,\theta)$. Set
\be
\label{e17}
 c_{++}  = 0
\ee
in Eq.(15). The remainder will possess an exceptionally rich structure. The
whole situation can be subdivided into three cases
\be
\label{e18}
 c_{+2} \neq 0 \ , \  c_{+ \overline{1}}\ =\ 0 \ (case\ I)
\ee

\be
\label{e19}
 c_{+2}\ =\ 0 \ , \  c_{+ \overline{1}} \neq 0 \ (case \ II)
\ee

\be
\label{e20}
 c_{+2}\ =\ 0 \ , \  c_{+ \overline{1}}\ =\ 0 \ (case \ III)
\ee
We emphasize that we keep the parameter $n$ in the representation (3),(4)
as a fixed, non-negative integer.

{\bf Case I.} The conditions (17) and (18) are fulfilled (see Fig.I).

{\it Case I.1.1.}  If
\[ (n+2) c_{+0} + n c_{+J} + 2 c_{+} = 0\ ,\]
\[ c_{+\overline{2}}=c_{\overline{1}}=0\ , \]
\be
\label{e21}
(n+1) c_{0\overline{2}}+2 c_{\overline{2}}=0\ ,
\ee
 then ${\bf T}_2(x,\theta)$ preserves ${\cal P}_{n,n-1}$ and
{\bf ${\cal P}_{n+1,n-1}$}.

{\it Case I.1.2.}  If
\[ (n+4+2m) c_{+0} + n c_{+J} + 2 c_{+} = 0\ ,\]
\[ c_{+\overline{2}}= c_{\overline{1}}=0\ , \]
\be
\label{e22}
 c_{0\overline{2}}= c_{\overline{2}}= c_{-\overline{2}} = 0\ ,
\ee
at a certain integer $m \geq 0$ , then ${\bf T}_2(x,\theta)$ preserves
${\cal P}_{n,n-1}$ and {\bf ${\cal P}_{n+2+m,n-1}$}. If $m$ is non-integer,
then ${\bf T}_2(x,\theta)$
preserves ${\cal P}_{n,n-1}$ and {\bf ${\cal P}_{\infty,n-1}$}.

{\it Case I.1.3.}  If
\[ (n+1) c_{+J} + 2 c_{+} = 0\ ,\]
\[ c_{+0}=0\ , \]
\[ c_{+\overline{2}}= c_{\overline{1}}=0\ , \]
\be
\label{e23}
 c_{0\overline{2}}= c_{\overline{2}}= c_{-\overline{2}} = 0\ ,
\ee
 then ${\bf T}_2(x,\theta)$ preserves the infinite flag
of polynomial spaces the {\bf ${\cal P}_{n+m,n-1},\ m=0,1,2,\ldots$}.

{\it Case I.2.1.}  If
\[ (n-3) c_{+0} + (n+1) c_{+J} + 2 c_{+} = 0\ ,\]
\[ (n-1)c_{+\overline{2}}=c_{\overline{1}}\ , \]
\be
\label{e24}
(n-1) c_{0\overline{2}}+2 c_{\overline{2}}=0\ ,
\ee
 then ${\bf T}_2(x,\theta)$ preserves ${\cal P}_{n,n-1}$ and
{\bf ${\cal P}_{n,n-2}$}.

{\it Case I.2.2.}  If
\[ 3 c_{+0} - c_{+J}  = 0\ ,\]
\[ (2k+2n+4) c_{+0} + 2 c_{+} = 0\ ,\]
\[ c_{+\overline{2}}=c_{\overline{1}}=0\ , \]
\be
\label{e25}
(2k-n+3) c_{0\overline{2}}+2 c_{\overline{2}}=0\ ,
\ee
at a certain integer $k \geq 0$, then ${\bf T}_2(x,\theta)$ preserves
${\cal P}_{n,n-1}$ and {\bf ${\cal P}_{k+2,k}$}.

{\it Case I.2.3.}  If
\[ c_{+0} = c_{+J}= c_{+} = 0\ ,\]
\[ c_{+\overline{2}}=c_{\overline{1}}=0\ , \]
\be
\label{e26}
 c_{0\overline{2}}= c_{\overline{2}}=0\ ,
\ee
 then ${\bf T}_2(x,\theta)$ preserves ${\cal P}_{n,n-1}$ and the infinite
flag of the polynomial spaces {\bf ${\cal P}_{k+2,k}, \ k=0,1,2,\ldots$}.
Note in general for this case $ c_{-\overline{2}} \neq 0$.

{\it Case I.3.1.}  If
\[ (n-5-2m) c_{+0} + (n+1) c_{+J} + 2 c_{+} = 0\ ,\]
\[ c_{+\overline{2}}= c_{\overline{1}}=0\ , \]
\be
\label{e27}
 c_{0\overline{2}}= c_{\overline{2}}= c_{-\overline{2}} = 0\ ,
\ee
at a certain integer $0 \leq m \leq (n-3)$ , then ${\bf T}_2(x,\theta)$
preserves ${\cal P}_{n,n-1}$ and {\bf ${\cal P}_{n,n-3-m}$}.

{\it Case I.3.2.}  If
\[ c_{+0}=0\ , \]
\[ (n+1) c_{+J} + 2 c_{+} = 0\ ,\]
\[ c_{+\overline{2}}= c_{\overline{1}}=0\ , \]
\be
\label{e28}
 c_{0\overline{2}}= c_{\overline{2}}= c_{-\overline{2}} = 0\ ,
\ee
 then ${\bf T}_2(x,\theta)$ preserves ${\cal P}_{n,n-1}$ and the sequence
of the polynomial spaces {\bf ${\cal P}_{n,n-3-m},\ m=0,1,2,\ldots , (n-3)$}.

{\it Case I.3.3.}  If
\[ (2k+1-n) c_{+0} + (n+1) c_{+J} + 2 c_{+} = 0\ ,\]
\[ (2m+5) c_{+0} - c_{+J} = 0\ ,\]
\[ c_{+\overline{2}}= c_{\overline{1}}=0\ , \]
\be
\label{e29}
 c_{0\overline{2}}= c_{\overline{2}}= c_{-\overline{2}} = 0\ ,
\ee
at  certain integers $k , m \geq 0 $ , then ${\bf T}_2(x,\theta)$ preserves
${\cal P}_{n,n-1}$ and {\bf ${\cal P}_{k+3+m,k}$}.

{\it Case I.3.4.}  If
\[  c_{+0} = c_{+J}= c_{+} = 0\ ,\]
\[ c_{+\overline{2}}= c_{\overline{1}}=0\ , \]
\be
\label{e30}
 c_{0\overline{2}}= c_{\overline{2}}= c_{-\overline{2}} = 0\ ,
\ee
 then ${\bf T}_2(x,\theta)$ preserves
${\cal P}_{n,n-1}$ and the infinite flag of polynomial spaces {\bf ${\cal
P}_{k+3+m,k}\ , k,m=0,1,2,\ldots$} (cf. {\it Cases I.1.3 and I.2.3}).

{\bf Case II.} The conditions (17) and (19) are fulfilled (see Fig.II).

{\it Case II.1.1.}  If
\[ (n+1) c_{+0} + (n+1) c_{+J} + 2 c_{+} = 0\ ,\]
\be
\label{e31}
 c_{2}= 0 ,
\ee
 then ${\bf T}_2(x,\theta)$ preserves ${\cal P}_{n,n-1}$ and
{\bf ${\cal P}_{n,n}$}.

{\it Case II.1.2.}  If
\[ (n+3) c_{+0} + (n+1) c_{+J} + 2 c_{+} = 0\ ,\]
\[(n+2) c_{01}+ 2c_{1}=0\ , \]
\be
\label{e32}
 c_{+1} = c_{2} = 0\ ,
\ee
  then ${\bf T}_2(x,\theta)$ preserves ${\cal P}_{n,n-1}$ and
{\bf ${\cal P}_{n,n+1}$}.

{\it Case II.1.3.}  If
\[ (2k+5+n) c_{+0} + (n+1) c_{+J} + 2 c_{+} = 0\ ,\]
\[  c_{01} = c_{1} = 0\ ,\]
\[ c_{+1} = c_{2} = 0\ , \]
\be
\label{e33}
 c_{-1} = 0\ ,
\ee
at a certain integer $k \geq 0$ , then ${\bf T}_2(x,\theta)$ preserves
${\cal P}_{n,n-1}$ and {\bf ${\cal P}_{n,n+2+k}$}.

{\it Case II.1.4.}  If
\[  c_{+0} = 0 , \]
\[ (n+1) c_{+J}+2 c_{+} = 0\ ,\]
\[  c_{01} = c_{1} = 0\ ,\]
\[ c_{+1} = c_{2} = 0\ , \]
\be
\label{e34}
 c_{-1} = 0 ,
\ee
 then ${\bf T}_2(x,\theta)$ preserves
${\cal P}_{n,n-1}$ and the infinite flag of polynomial spaces {\bf ${\cal
P}_{n,n+k} , k = 0,1,2,\ldots$}
 (cf. {\it Cases I.1.3 , I.2.3 and I.3.4}).

{\it Case II.2.1.}  If
\[ (n-2) c_{+0} + n c_{+J} + 2 c_{+} = 0\ ,\]
\be
\label{e35}
 c_{+1} = c_{2} \ ,
\ee
then ${\bf T}_2(x,\theta)$ preserves ${\cal P}_{n,n-1}$ and
{\bf ${\cal P}_{n-1,n-1}$}.

{\it Case II.2.2.}  If
\[ (n+1) c_{+0} -  c_{+} = 0\ ,\]
\[ 3 c_{+0} +  c_{+J}  = 0\ ,\]
\[ n c_{01}+ 2c_{1}=0\ , \]
\be
\label{e36}
 c_{+1} = c_{2}=0\ ,
\ee
then ${\bf T}_2(x,\theta)$ preserves ${\cal P}_{n,n-1}$ and
{\bf ${\cal P}_{n-1,n}$}.

{\it Case II.2.3.}  If
\[ (n-2) c_{+0} + n c_{+J} + 2 c_{+} = 0\ ,\]
\[ (2k+1) c_{+0} +  c_{+J} = 0\ ,\]
\[ c_{+1} = c_{2} = 0\ , \]
\be
\label{e37}
 c_{01} = c_{1} = c_{-1}= 0\ ,
\ee
at a  certain integer $k \geq 0$,  then ${\bf T}_2(x,\theta)$
preserves ${\cal P}_{n,n-1}$ and {\bf ${\cal P}_{n-1,n+k+1}$}.

{\it Case II.2.4.}  If
\[  c_{+0} = c_{+J}= c_{+} = 0\ ,\]
\[ c_{+1} = c_{2} = 0\ , \]
\be
\label{e38}
 c_{01} = c_{1} = c_{-1}= 0\ ,
\ee
 then ${\bf T}_2(x,\theta)$ preserves
${\cal P}_{n,n-1}$ and the infinite flag of the polynomial spaces
{\bf ${\cal P}_{n-1,n+k} , k = 0,1,2,\ldots$}
 (cf. {\it Cases I.1.3 , I.2.3 , I.3.4 and II.1.4}).

{\it Case II.3.1.}  If
\[ (n-4)c_{+0}  + nc_{+J} + 2c_{+} = 0\ ,\]
\[ (n-2) c_{01}+ 2c_{1}=0\ , \]
\be
\label{e39}
 c_{+1} = c_{2} = 0\ ,
\ee
 then ${\bf T}_2(x,\theta)$ preserves ${\cal P}_{n,n-1}$ and
{\bf ${\cal P}_{n-2,n-1}$}.

{\it Case II.4.1.}  If
\[ (m-2n) c_{+0} +  c_{+} = 0\ ,\]
\[ 3 c_{+0} +  c_{+J}  = 0\ ,\]
\[ (2m+2-n) c_{01}+ 2c_{1}=0\ , \]
\be
\label{e40}
 c_{+1} = c_{2} = 0\ ,
\ee
at a  certain integer $m \geq 0$, then ${\bf T}_2(x,\theta)$ preserves
${\cal P}_{n,n-1}$ and {\bf ${\cal P}_{m,m+1}$}.

{\it Case II.4.2.}  If
\[ (2m-n)c_{+0}  + nc_{+J} + 2c_{+} = 0\ ,\]
\[ (2k+5) c_{+0} + 2 c_{+J}  = 0\ ,\]
\[ c_{+1} = c_{2}=0 , \]
\be
\label{e41}
 c_{01} = c_{1} = c_{-1}= 0
\ee
at  certain integers $k \geq 0 , m \geq 0$, then ${\bf T}_2(x,\theta)$
preserves ${\cal P}_{n,n-1}$ and {\bf ${\cal P}_{m,m+2+k}$}.

{\it Case II.4.3.}  If
\[ c_{+0}  = c_{+J} = c_{+} = 0\ ,\]
\[ c_{+1} = c_{2}=0 , \]
\be
\label{e42}
 c_{01} = c_{1} = c_{-1}= 0
\ee
 then ${\bf T}_2(x,\theta)$ preserves ${\cal P}_{n,n-1}$ and the infinite
flag of the polynomial spaces {\bf ${\cal P}_{m,m+1+k},\ m,k=0,1,2,\ldots $}
 (cf. {\it Cases I.1.3 , I.2.3 , I.3.4 , II.1.4 and II.2.4}).

{\bf Case III.} The conditions (17) and (20) are fulfilled (see Fig.III).

{\it Case III.1.1.}  If
\[ (2m-n)c_{+0}  + nc_{+J} + 2c_{+} = 0\ ,\]
\[  c_{+0} +  c_{+J}  = 0\ ,\]
\be
\label{e43}
 (m-n) c_{+1} + c_{2}=0 ,
\ee
at a certain integer $ m \geq 0$, then ${\bf T}_2(x,\theta)$ preserves
${\cal P}_{n,n-1}$ and {\bf ${\cal P}_{m,m}$}.

{\it Case III.1.2.}  If
\[ c_{+0}  = c_{+J} = c_{+} = 0\ ,\]
\be
\label{e44}
 c_{+1} = c_{2}=0 ,
\ee
 then ${\bf T}_2(x,\theta)$ preserves ${\cal P}_{n,n-1}$ and the infinite
flag of the polynomial spaces {\bf ${\cal P}_{m,m},\ m=0,1,2,\ldots  $}
 (cf. {\it Cases I.1.3 , I.2.3 , I.3.4 , II.1.4, II.2.4 and
 II.4.3}).

{\it Case III.2.1.}  If
\[ (2m-n)c_{+0}  + nc_{+J} + 2c_{+} = 0\ ,\]
\[  c_{+0} - c_{+J}  = 0\ ,\]
\be
\label{e45}
 m c_{+ \overline{2}} - c_{\overline{1}}=0 ,
\ee
at a certain integer $ m \geq 0$, then ${\bf T}_2(x,\theta)$ preserves
${\cal P}_{n,n-1}$ and {\bf ${\cal P}_{m,m-1}$}.

{\it Case III.2.2.}  If
\[ c_{+0}  = c_{+J} = c_{+} = 0\ ,\]
\be
\label{e46}
 c_{+ \overline{2}} = c_{\overline{1}}=0
\ee
 then ${\bf T}_2(x,\theta)$ preserves ${\cal P}_{n,n-1}$ and the infinite
flag of polynomial spaces {\bf ${\cal P}_{m,m-1},\ m=0,1,2,\ldots  $}
 (cf. {\it Cases I.1.3 , I.2.3 , I.3.4 , II.1.4, II.2.4,
II.4.3 and III.1.2}). This case corresponds to exactly-solvable operators
${\bf E}_k$.

In \cite{t2} it has been shown that under a certain condition some
quasi-exactly-solvable operators $T_2(x)$ in one real
variable can preserve
two polynomial spaces of different dimensions $n$ and $m$.
It has been shown that
those quasi-exactly-solvable operators  $T_2(x)$ can be represented through
the generators of $sl_2({\bf R})$ in projectivized representation
characterized either by the mark $n$ or by the mark $m$.
The above analysis shows that the quasi-exactly-solvable operators
${\bf T}_2(x,\theta)$
in two variables (one real and one Grassmann) possess an extremely rich
variety of internal properties . They are characterized by different
structures of invariant sub-spaces. However, generically the
quasi-exactly-solvable operators  ${\bf T}_2(x,\theta)$ can preserve
either one, or two,
or infinitely many polynomial spaces. For the latter, those operators become
'exactly-solvable' (see  {\it Cases I.1.3 , I.2.3 , I.3.4 ,
 II.1.4, II.2.4, II.4.3 and III.1.2})
  giving rise to the eigenvalue problems (8) possessing
infinite sequences of polynomial eigenfunctions. In general, for the two
latter cases the interpretation of  ${\bf T}_2(x,\theta)$
in term of $osp(2,2)$ generators characterized by different marks
 does not exist unlike the case of quasi-exactly-solvable operators
in one real variable. The only exceptions are given  by the
{\it Case III.2.1} and {\it Case III.2.2}.

\section{2 x 2 matrix differential equations in $x$}

\hspace{13mm}
It is well-known that anti-commuting variables can be represented by
matrices.
In our case the matrix representation is as follows: substitute $\theta$ and
$\partial_{\theta}$ in the generators (3),(4) by the Pauli matrices $\sigma^+$
and $\sigma^-$, respectively, acting on two-component spinors. In fact, all
main notations are preserved like quasi-exactly-solvable and exactly-solvable
operator, grading etc.

In the explicit form the fermionic generators (4) in matrix representation are
written as follows:
\be
\label{e47}
Q \ = \ { \sigma^- \brack x\sigma^-}\ , \
{\bar Q} \ = \ { x\sigma^+ \partial_x-n\sigma^+ \brack
-\sigma^-\partial_x} .
\ee
The representation (47) implies that in the spectral problem (8) an
eigenfunction $\varphi(x)$ is treated  as a two-component spinor
\be
\label{e48}
\varphi(x) \ = \ { \varphi_1(x) \brack \varphi_2(x)} ,
\ee
In the matrix formalism, the polynomial space (1) has the form:
\be
\label{e49}
{\cal P}_{N,M} \ = \ \left \langle  \begin{array}{c}
x^0,x^1,\dots,x^M  \\
x^0, x^1, \dots,x^N \end{array} \right \rangle
\ee
where the terms of zero degree in $\theta$ come in as the lower component and
the terms of first degree in $\theta$ come in as the upper component.
The operator  ${\bf T}_k (x,\theta)$ becomes a 2x2 matrix differential
operator ${\bf T}_k (x)$ having
derivatives in $x$ up to $k$-th order. In order to distinguish the matrix
operator in $x$ from the operator in $x,\theta$, we will denote the former as
${\bf T}_k (x)$.
Finally, as a consequence of Theorem 1,
 we arrive at the eigenvalue problem for a 2x2 matrix quasi-exactly-solvable
differential operator ${\bf T}_k (x)$, possessing in general $(2n+1)$
polynomial solutions of the form ${\cal P}_{N,N-1}$. This eigenvalue
problem can be written in the form (cf.Eq.(9))
\be
\label{e50}
 \sum_{i=0}^{i=k} {\bf a}_{k,i} (x) \varphi^{(i)}_{x}
(x) \ = \ \varepsilon \varphi(x) \ ,
\ee
 where the notation $ \varphi^{(i)}_{x}$ means the
$i-$th order derivative with respect to $x$ of each component of the spinor
$\varphi(x)$ (see Eq.(48)). The coefficient functions $ {\bf a}_{k,i} (x)$
are 2x2  matrices and generically for the $k$-th order
quasi-exactly-solvable operator their matrix elements are polynomials.
Suppose that $k>0$. Then the matrix elements are given
by the following expressions
\be
\label{e51}
  {\bf a}_{k,i} (x) \ = \ \left( \begin{array}{cc}
A_{k,i}^{[k+i]} & B_{k,i}^{[k+i-1]} \\
C_{k,i}^{[k+i+1]} & D_{k,i}^{[k+i]}
\end{array}  \right)
\ee
at $k > 0$, where the superscript in square brackets displays the order of
the corresponding polynomial.

 It is easy to calculate the number of free parameters of a
quasi-exactly-solvable operator ${\bf T}_k (x)$
\be
\label{e52}
par({\bf T}_k (x))= 4 (k+1)^2
\ee
(cf.Eq.(12)).

For the case of exactly-solvable problems, the matrix elements (53)
of the coefficient functions will be modified
\be
\label{e53}
  {\bf a}_{k,i} (x) \ = \ \left( \begin{array}{cc}
A_{k,i}^{[i]} & B_{k,i}^{[i-1]} \\
C_{k,i}^{[i+1]} & D_{k,i}^{[i]}
\end{array}  \right)
\ee
where $k > 0$. An infinite family of orthogonal polynomials as
eigenfunctions of Eq.(50), if they exist, will occur, if and only
if the coefficients functions have the form (53). The number of free
parameters of an exactly-solvable operator ${\bf E}_k (x)$ and,
correspondingly, the maximal number of free parameters of the 2 x 2
matrix orthogonal polynomials in one real variable is equal to
\be
\label{e54}
par({\bf E}_k (x))= 2k (k+3) + 3
\ee
(cf.Eq.(14)).

The increase in the number of free parameters for the 2 x 2 matrix
operators with respect to the case of the operators in $x, \theta$ is
connected to the occurance of extra monomials of degree $(k+1)$ in generators
of $osp(2,2)$ (see Eqs.(3),(4),(47)), leading to the $k$-th order differential
operators in $x$.

 Thus, the above formulas describe the coefficient functions of matrix
differential equations (50), which can possess polynomials in $x$ as
solutions,  resolving the analogue of the
generalized Bochner problem Eq.(0) for the case of 2 x 2 matrix differential
equations in one real variable.

Now let us take the quasi-exactly-solvable matrix operator ${\bf T}_2 (x)$
 and try to reduce Eq.(8) to the Schroedinger equation
\be
\label{55}
[ -{1 \over 2} {d^2 \over dy^2} + \hat{V}(y) ] \Psi (y)\ =\ E \Psi (y)
\ee
where $ \hat{V}(y)$ is a two-by-two {\it hermitian} matrix, by making a
change of variable $x \mapsto y$ and "gauge" transformation
\be
\label{e56}
\Psi \ = \ U \varphi
\ee
where $U$ is an arbitrary two-by-two matrix depending on
the variable $y$. In order to get some "reasonable" Schroedinger equation one
should fulfill two requirements: (i) the  potential  $ \hat{V}(y)$ must be
hermitian and (ii) the eigenfunctions $\Psi(y)$ must belong to a
certain Hilbert space.

Unlike the case of quasi-exactly-solvable differential operators in one real
variable (see \cite{olver}), this problem has no complete solution so far.
Therefore it seems instructive to display a particular example \cite{st}.

Consider the quasi-exactly-solvable operator
\[ {\bf T}_2 = - 2 T^0 T^- + 2 T^- J - i \beta T^0 Q_1 + \]
\be
\label{e57}
\alpha T^0 - (2n+1) T^- - {i\beta \over 2}(3n + 1) Q^1 + {i \over 2}\alpha
\beta Q^2 - i \beta \overline{Q}_1 \ ,
\ee
where $\alpha$ and $\beta$ are parameters. Upon introducing a new
variable $y=x^2$ and  after straightforward calculations
one finds the following expression for the matrix $U$ in Eq.(56)
\be
\label{e58}
U = \exp ( - {\alpha y^2 \over 4} + {i \beta y^2 \over 4} \sigma_1)
\ee
and for the potential $\hat{V}$ in Eq.(56)
\[ \hat{V} (y) = {1 \over 8} (\alpha ^2 - \beta ^2) y^2 + \sigma _2
[-(n + {1 \over 4}) \beta + {\alpha \beta \over 4} y^2 - {\alpha \over 4}
\tan {\beta y^2 \over 2}] \cos {\beta y^2 \over 2} +  \]
\be
\label{e59}
\sigma _3 [-(n + {1 \over 4}) \beta + {\alpha \beta \over 4} y^2 -
{\alpha \over 4}
\cot {\beta y^2 \over 2}] \sin {\beta y^2 \over 2}
\ee
It is easy to see that the potential $\hat{V}$ is hermitian;
$(2n+1)$ eigenfunctions have the form of polynomials multiplied
by the exponential factor $U$ and they are obviously normalizable.

At closing, I am very indebted to V.Arnold for numerous useful
discussions and also to J.Frohlich and M.Shubin for  interest
in the subject and valuable suggestions.
Also I am very grateful to the Centre de Physique Theorique , where
this work was initiated, and to the Institute of
Theoretical Physics, ETH-Honggerberg, where this work was completed,
for their kind hospitality extended to me.

\newpage
\begin{center}
{\bf FIGURE CAPTIONS}
\end{center}

Figs. 0--III .

Newton diagrams describing invariant subspaces ${\cal P}_{N,M}$
 of the second-order polynomials in the generators of $osp(2,2)$.
The lower line corresponds to the part of the space of zero
degree in  $\theta$ and the upper line of first degree in $\theta$.
The letters without brackets indicate the maximal degree of the polynomial
in $x$. The letters in brackets indicate the maximal (or minimal) possible
degree, if the degree can be varied. The {\it thin} line displays
schematically the length of polynomial in $x$ (the number of monomials).
The {\it thick} line shows, that the length of polynomial can not be more
(or less) than that size. The {\it dashed} line means, that the length of
polynomial can take any size on this line. If the dashed line is unbounded,
it means that the degree of polynomial can be arbitrary up to infinity.
Numbering the figures I-III corresponds to the cases, which satisfy the
conditions (17),(18) ({\it Case I}); (17),(19) ({\it Case II}) and (17),(20)
({\it Case III}).


\newpage
\noindent
\begin{picture}(400,50)(-10,-20)
\linethickness{1.2pt}
\put(10,10){\circle*{5}}
\put(10,10){\line(1,0){50}}
\put(60,10){\circle*{5}}
\put(10,20){\circle*{5}}
\put(10,20){\line(1,0){40}}
\put(50,20){\circle*{5}}
\put(-10,7){$x$}
\put(-10,17){$\Theta$}
\put(35,25){$n-1$}
\put(60,0){$n$}
\end{picture}
\begin{center}
Fig. 0. \ Basic subspace
\end{center}
\begin{picture}(400,50)(-10,-20)
\linethickness{1.2pt}
\put(10,10){\circle*{5}}
\put(10,10){\line(1,0){50}}
\put(60,10){\circle*{5}}
\put(10,20){\circle*{5}}
\put(10,20){\line(1,0){30}}
\put(40,20){\circle*{5}}
\put(-10,7){$x$}
\put(-10,17){$\Theta$}
\put(25,25){$n-1$}
\put(50,-2){$n+1$}
\put(22,-20){(a)}
\put(130,10){\circle*{5}}
\linethickness{2.4pt}
\put(130,10){\line(1,0){40}}
\put(170,10){\circle*{5}}
\linethickness{1.2pt}
\put(170,10){\line(1,0){30}}
\put(200,10){\circle*{5}}
\put(130,20){\circle*{5}}
\put(130,20){\line(1,0){30}}
\put(160,20){\circle*{5}}
\put(150,25){$n-1$}
\put(148,-2){$(n+2)$}
\put(192,-2){$n+2+m$}
\put(182,-20){(b)}
\linethickness{2.4pt}
\put(290,10){\circle*{5}}
\put(290,10){\line(1,0){40}}
\put(330,10){\circle*{5}}
\linethickness{1.2pt}
\put(290,20){\circle*{5}}
\put(290,20){\line(1,0){30}}
\put(320,20){\circle*{5}}
\put(320,10){\dashbox{2}(40,0){}}
\put(310,25){$n-1$}
\put(315,-2){$(n+2)$}
\put(320,-20){(c)}
\end{picture}
\begin{center}
Fig. I.1 . \  Subspaces for the {\it Case I.1}
\end{center}
\begin{picture}(400,50)(-10,-20)
\linethickness{1.2pt}
\put(10,10){\circle*{5}}
\put(10,10){\line(1,0){50}}
\put(60,10){\circle*{5}}
\put(10,20){\circle*{5}}
\put(10,20){\line(1,0){25}}
\put(35,20){\circle*{5}}
\put(-10,7){$x$}
\put(-10,17){$\Theta$}
\put(25,25){$n-2$}
\put(57,-2){$n$}
\put(22,-20){(a)}
\put(130,10){\circle*{5}}
\linethickness{2.4pt}
\put(130,10){\line(1,0){10}}
\put(140,10){\circle*{5}}
\linethickness{1.2pt}
\put(140,10){\line(1,0){60}}
\put(200,10){\circle*{5}}
\put(130,20){\circle*{5}}
\put(130,20){\line(1,0){30}}
\put(160,20){\circle*{5}}
\put(157,25){$k$}
\put(133,-3){$(2)$}
\put(192,-2){$k+2$}
\put(182,-20){(b)}
\linethickness{2.4pt}
\put(290,10){\circle*{5}}
\put(290,10){\line(1,0){10}}
\put(300,10){\circle*{5}}
\linethickness{1.2pt}
\put(290,20){\circle*{5}}
\put(290,20){\dashbox{2}(40,0){}}
\put(300,10){\dashbox{2}(60,0){}}
\put(295,-3){$(2)$}
\put(327,25){$k$}
\put(352,-2){$k+2$}
\put(320,-20){(c)}
\end{picture}
\begin{center}
Fig. I.2 . \  Subspaces for the {\it Case I.2}
\end{center}
\begin{picture}(400,50)(-10,-20)
\linethickness{1.2pt}
\put(10,10){\circle*{5}}
\put(10,10){\line(1,0){50}}
\put(60,10){\circle*{5}}
\linethickness{2.4pt}
\put(10,20){\circle*{5}}
\put(10,20){\line(1,0){40}}
\put(25,20){\circle*{7}}
\put(5,28){$n-3+m$}
\put(50,20){\circle*{5}}
\put(-10,7){$x$}
\put(-10,17){$\Theta$}
\put(60,21){$(n-3)$}
\put(60,0){$n$}
\put(22,-20){(a)}
\linethickness{1.2pt}
\put(120,10){\circle*{5}}
\put(120,10){\line(1,0){60}}
\put(180,10){\circle*{5}}
\put(120,20){\circle*{5}}
\put(120,20){\dashbox{2}(40,0){}}
\put(160,20){\circle*{5}}
\put(165,21){$(n-3)$}
\put(180,0){$n$}
\put(142,-20){(b)}
\put(250,10){\circle*{5}}
\linethickness{2.4pt}
\put(250,10){\line(1,0){12}}
\put(262,10){\circle*{5}}
\linethickness{1.2pt}
\put(262,10){\line(1,0){50}}
\put(312,10){\circle*{5}}
\put(250,20){\circle*{5}}
\put(250,20){\line(1,0){30}}
\put(280,20){\circle*{5}}
\put(277,25){$k$}
\put(256,-3){$(3)$}
\put(282,-2){$k+3+m$}
\put(280,-20){(c)}
\put(380,10){\circle*{5}}
\linethickness{2.4pt}
\put(380,10){\line(1,0){12}}
\put(392,10){\circle*{5}}
\linethickness{1.2pt}
\put(392,10){\dashbox{2}(50,0){}}
\put(380,20){\circle*{5}}
\put(380,20){\dashbox{2}(30,0){}}
\put(407,25){$k$}
\put(386,-3){$(3)$}
\put(412,-2){$k+3+m$}
\put(410,-20){(d)}
\end{picture}
\begin{center}
Fig. I.3 . \  Subspaces for the {\it Case I.3}
\end{center}
\begin{picture}(400,50)(-10,-20)
\linethickness{1.2pt}
\put(10,10){\circle*{5}}
\put(10,10){\line(1,0){30}}
\put(40,10){\circle*{5}}
\put(10,20){\circle*{5}}
\put(10,20){\line(1,0){30}}
\put(40,20){\circle*{5}}
\put(-10,7){$x$}
\put(-10,17){$\Theta$}
\put(45,25){$n$}
\put(45,0){$n$}
\put(20,-20){(a)}

\put(110,10){\circle*{5}}
\put(110,10){\line(1,0){30}}
\put(140,10){\circle*{5}}
\put(110,20){\circle*{5}}
\put(110,20){\line(1,0){40}}
\put(150,20){\circle*{5}}
\put(145,25){$n+1$}
\put(145,0){$n$}
\put(120,-20){(b)}

\put(230,10){\circle*{5}}
\put(230,20){\circle*{5}}
\linethickness{2.4pt}
\put(230,20){\line(1,0){40}}
\put(270,20){\circle*{5}}
\linethickness{1.2pt}
\put(270,20){\line(1,0){20}}
\put(290,20){\circle*{5}}
\put(230,10){\line(1,0){25}}
\put(255,10){\circle*{5}}
\put(245,28){$(n+2)$}
\put(289,25){$n+2+k$}
\put(250,-2){$n-1$}
\put(252,-20){(c)}
\put(380,10){\circle*{5}}
\put(380,20){\circle*{5}}
\linethickness{2.4pt}
\put(380,20){\line(1,0){30}}
\put(410,20){\circle*{5}}
\linethickness{1.2pt}
\put(410,20){\dashbox{2}(35,0){}}
\put(380,10){\line(1,0){30}}
\put(410,10){\circle*{5}}
\put(405,28){$(n)$}
\put(441,25){$n+k$}
\put(408,-2){$n$}
\put(410,-20){(d)}
\end{picture}
\begin{center}
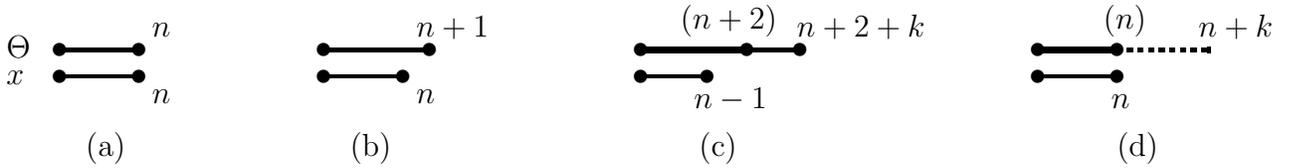
Fig. II.1 . \ Subspaces for the {\it Case II.1}
\end{center}
\newpage
\noindent
\begin{picture}(400,50)(-10,-20)
\linethickness{1.2pt}
\put(10,10){\circle*{5}}
\put(10,10){\line(1,0){30}}
\put(40,10){\circle*{5}}
\put(10,20){\circle*{5}}
\put(10,20){\line(1,0){30}}
\put(40,20){\circle*{5}}
\put(-10,7){$x$}
\put(-10,17){$\Theta$}
\put(43,25){$n-1$}
\put(43,0){$n-1$}
\put(20,-20){(a)}
\put(110,10){\circle*{5}}
\put(110,10){\line(1,0){30}}
\put(140,10){\circle*{5}}
\put(110,20){\circle*{5}}
\put(110,20){\line(1,0){40}}
\put(150,20){\circle*{5}}
\put(145,25){$n$}
\put(138,0){$n-1$}
\put(120,-20){(b)}
\put(230,10){\circle*{5}}
\put(230,20){\circle*{5}}
\linethickness{2.4pt}
\put(230,20){\line(1,0){35}}
\put(265,20){\circle*{5}}
\linethickness{1.2pt}
\put(265,20){\line(1,0){35}}
\put(300,20){\circle*{5}}
\put(230,10){\line(1,0){25}}
\put(255,10){\circle*{5}}
\put(245,28){$(n+1)$}
\put(289,25){$n+1+k$}
\put(250,-2){$n-1$}
\put(252,-20){(c)}
\put(380,10){\circle*{5}}
\put(380,20){\circle*{5}}
\linethickness{2.4pt}
\put(380,20){\line(1,0){35}}
\put(415,20){\circle*{5}}
\linethickness{1.2pt}
\put(415,20){\dashbox{2}(35,0){}}
\put(380,10){\line(1,0){25}}
\put(405,10){\circle*{5}}
\put(405,28){$(n)$}
\put(441,25){$n+k$}
\put(408,-2){$n-1$}
\put(410,-20){(d)}
\end{picture}
\begin{center}
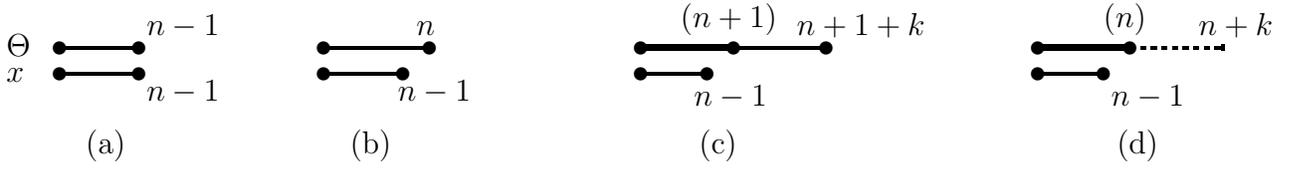
Fig. II.2 . \ Subspaces for the {\it Case II.2}
\end{center}
\begin{picture}(400,50)(-10,-20)
\linethickness{1.2pt}
\put(10,10){\circle*{5}}
\put(10,10){\line(1,0){30}}
\put(40,10){\circle*{5}}
\put(10,20){\circle*{5}}
\put(10,20){\line(1,0){40}}
\put(50,20){\circle*{5}}
\put(-10,7){$x$}
\put(-10,17){$\Theta$}
\put(53,25){$n-1$}
\put(43,0){$n-2$}
\end{picture}
\begin{center}
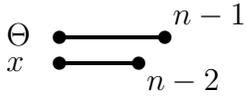
Fig. II.3 . \ Subspaces for the {\it Case II.3}
\end{center}
\begin{picture}(400,50)(-10,-20)
\put(10,20){\circle*{5}}
\linethickness{2.4pt}
\put(10,20){\line(1,0){10}}
\put(20,20){\circle*{5}}
\linethickness{1.2pt}
\put(20,20){\line(1,0){40}}
\put(60,20){\circle*{5}}
\put(10,10){\circle*{5}}
\put(10,10){\line(1,0){40}}
\put(50,10){\circle*{5}}
\put(57,25){$m+1$}
\put(16,28){$(1)$}
\put(52,-2){$m$}
\put(22,-20){(a)}

\put(130,20){\circle*{5}}
\linethickness{2.4pt}
\put(130,20){\line(1,0){20}}
\put(150,20){\circle*{5}}
\linethickness{1.2pt}
\put(150,20){\line(1,0){40}}
\put(190,20){\circle*{5}}
\put(130,10){\circle*{5}}
\put(130,10){\line(1,0){40}}
\put(170,10){\circle*{5}}
\put(185,25){$m+2+k$}
\put(145,28){$(2)$}
\put(170,-2){$m$}
\put(142,-20){(b)}
\put(250,20){\circle*{5}}
\linethickness{2.4pt}
\put(250,20){\line(1,0){20}}
\put(270,20){\circle*{5}}
\linethickness{1.2pt}
\put(270,20){\dashbox{2}(40,0){}}
\put(250,10){\circle*{5}}
\put(250,10){\dashbox{2}(40,0){}}
\put(305,25){$m+2+k$}
\put(265,28){$(2)$}
\put(290,-2){$m$}
\put(272,-20){(c)}
\end{picture}
\begin{center}
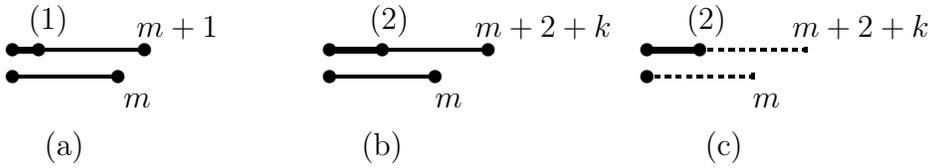
Fig. II.4 . \ Subspaces for the {\it Case II.4}
\end{center}
\begin{picture}(400,50)(-10,-20)
\linethickness{1.2pt}
\put(10,10){\circle*{5}}
\put(10,10){\line(1,0){30}}
\put(40,10){\circle*{5}}
\put(10,20){\circle*{5}}
\put(10,20){\line(1,0){30}}
\put(40,20){\circle*{5}}
\put(-10,7){$x$}
\put(-10,17){$\Theta$}
\put(43,25){$m$}
\put(43,0){$m$}
\put(20,-20){(a)}
\linethickness{1.2pt}
\put(120,10){\circle*{5}}
\put(120,10){\dashbox{2}(40,0){}}
\put(120,20){\circle*{5}}
\put(120,20){\dashbox{2}(40,0){}}
\put(163,25){$m$}
\put(163,0){$m$}
\put(130,-20){(b)}
\end{picture}
\begin{center}
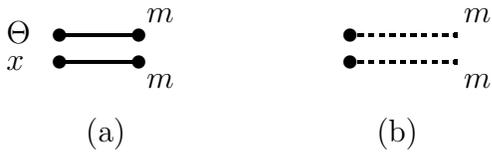
Fig. III.1 . \ Subspaces for the {\it Case III.1}
\end{center}
\begin{picture}(400,50)(-10,-20)
\put(10,10){\circle*{5}}
\linethickness{2.4pt}
\put(10,10){\line(1,0){10}}
\put(20,10){\circle*{5}}
\linethickness{1.2pt}
\put(20,10){\line(1,0){50}}
\put(70,10){\circle*{5}}
\put(10,20){\circle*{5}}
\put(10,20){\line(1,0){50}}
\put(60,20){\circle*{5}}
\put(62,25){$m-1$}
\put(17,-3){$(1)$}
\put(72,-2){$m$}
\put(42,-20){(a)}
\put(120,10){\circle*{5}}
\linethickness{2.4pt}
\put(120,10){\line(1,0){10}}
\put(130,10){\circle*{5}}
\linethickness{1.2pt}
\put(130,10){\dashbox{2}(50,0){}}
\put(120,20){\circle*{5}}
\put(120,20){\dashbox{2}(50,0){}}
\put(172,25){$m-1$}
\put(127,-3){$(1)$}
\put(182,-2){$m$}
\put(152,-20){(b)}
\end{picture}
\begin{center}
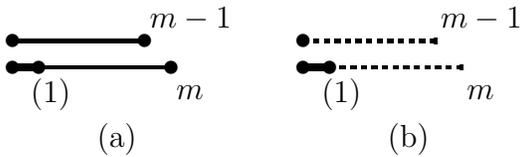
Fig. III.2 . \ Subspaces for the {\it Case III.2}
\end{center}



\newpage
\vfill
\begin{thebibliography}{299}
\bibitem{t1}
       "On polynomial solutions of differential equations"\\
       A.V.Turbiner, Preprint CPT-91/P.2628;\ J.Math.Phys.(in print)
\bibitem{t2}
       "Lie-algebraic approach to the theory of polynomial solutions.\\
       I. Ordinary differential equations and finite-difference equations in
       one variable"\\
       A.V.Turbiner, Preprint CPT-91/P.2679 (submitted to Comm.Math.Phys.)
\bibitem{st}
      "Quantal problems with partial algebraization of the spectrum"\\
       M.A. Shifman and A.V. Turbiner, {\it Comm. Math. Phys.} {\bf 120}
        347-365 (1989)
\bibitem{olver}
       "Normalizability properties of one-dimensional quasi-exactly-solvable
        Schroedinger operators "\\
         A. Gonzales-Lopez, N. Kamran and P.J. Olver. Preprint of the School
         of Math., Univ. Minnesota (Invited talk at the Session of AMS No.873,
         March 20-21, Springfield, USA)
\bibitem{w}
      "The theory of groups and Quantum mechanics"\\
       H.Weyl, Dover Publications, Inc. (1931)

\end{thebibliography}
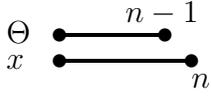
\end{document}